%% file: DLT_InteropAccess_Work.tex
\documentclass[10pt,journal]{IEEEtran}
\usepackage{amsmath,amssymb,amsfonts}
\usepackage{algorithmic}
\usepackage{graphicx}
\usepackage{textcomp}
\usepackage{xcolor}
\usepackage[nocompress]{cite}
\usepackage[ruled,vlined,linesnumbered]{algorithm2e}
\usepackage{tikz}
\usepackage{pifont}
\usepackage{pgfplots}
\pgfplotsset{compat=1.18}
\usepackage{xspace}
\usepackage{caption}
\usepackage{subcaption}
\usepackage{colortbl}
\usepackage{booktabs}
\usepackage{url}
\usepackage{amsmath}
\usepackage{paralist}
\usepackage{balance}

\def\BibTeX{{\rm B\kern-.05em{\sc i\kern-.025em b}\kern-.08em
    T\kern-.1667em\lower.7ex\hbox{E}\kern-.125emX}}

\newcommand{\ours}{\textit{InterAcct}\xspace}
\usepackage{enumitem}
\usepackage{mathtools}

\begin{document}


	%

\title{Cross-chain Access Control for Permissioned Blockchain Interoperation}

\author{Tirthankar~Sengupta,~\affiliation{IIT~Kharagpur,~India}~\texttt{mailtisen03@kgpian.iitkgp.ac.in}
        Bishakh~Chandra~Ghosh,~\affiliation{IIT~Kharagpur,~India}~\texttt{ghoshbishakh@gmail.com}
        Sandip~Chakraborty,~\affiliation{IIT~Kharagpur,~India}~\texttt{sandipc@cse.iitkgp.ac.in}
        and~Shamik~Sural,~~\affiliation{IIT~Kharagpur,~India}~\texttt{shamik@cse.iitkgp.ac.in}}%
\author{Tirthankar Sengupta,~Bishakh~Chandra~Ghosh,~Sandip~Chakraborty, and~Shamik~ Sural

\thanks{Tirthankar Sengupta (mailtisen03@kgpian.iitkgp.ac.in), Sandip Chakraborty (sandipc@cse.iitkgp.ac.in) , and Shamik Sural (shamik@cse.iitkgp.ac.in) are with the Department of Computer Science and
Engineering, Indian Institute of Technology, Kharagpur, India. Bishakh Chandra Ghosh (ghoshbishakh@gmail.com) is with Pinggy Technology Private Limited, India. }
}

 \maketitle 
\begin{abstract}
 \input{Main_Matter/abstract}
\end{abstract}



	\IEEEdisplaynontitleabstractindextext

	%
	\IEEEpeerreviewmaketitle


\section{Introduction}
\input{Main_Matter/introduction}
\section{Related Work}
\input{Main_Matter/relatedwork}
\section{Proposed Methodology}
\input{Main_Matter/Methodology}

\section{Security Analysis}
\input{Main_Matter/analysis}
\section{Implementation Details}
\input{Main_Matter/implementation}
\section{Experimental Results}
\input{Main_Matter/results}
\section{Conclusion and Future Directions}
\input{Main_Matter/conclusion_futurework}
\bibliographystyle{IEEEtran}
\bibliography{sample-base}
\end{document}

%% file: Main_Matter/abstract.tex
As enterprise blockchains become increasingly interconnected, access control must extend beyond the boundaries of a single network. Existing approaches mainly focus on controlling access within one blockchain, while cross-chain interactions involve multiple networks that may follow different access-control policies and administrative rules. In this paper, we propose \ours, an end-to-end access-control framework for secure interactions across permissioned blockchain networks. To avoid exposing sensitive internal information, \ours converts outgoing requests into a consortium-level representation before they are shared with another network. When a response is returned, the framework securely maps it back to the original requester within the source consortium, preserving confidentiality throughout the interaction. Our experimental results show that \ours can enforce end-to-end access control for cross-chain request-response interactions with low additional overhead and can continue to operate effectively as the workload increases.

%% file: Main_Matter/introduction.tex
\label{sec:intro}

\begin{figure}[!t]
\centering 
\includegraphics[width=0.8\linewidth]{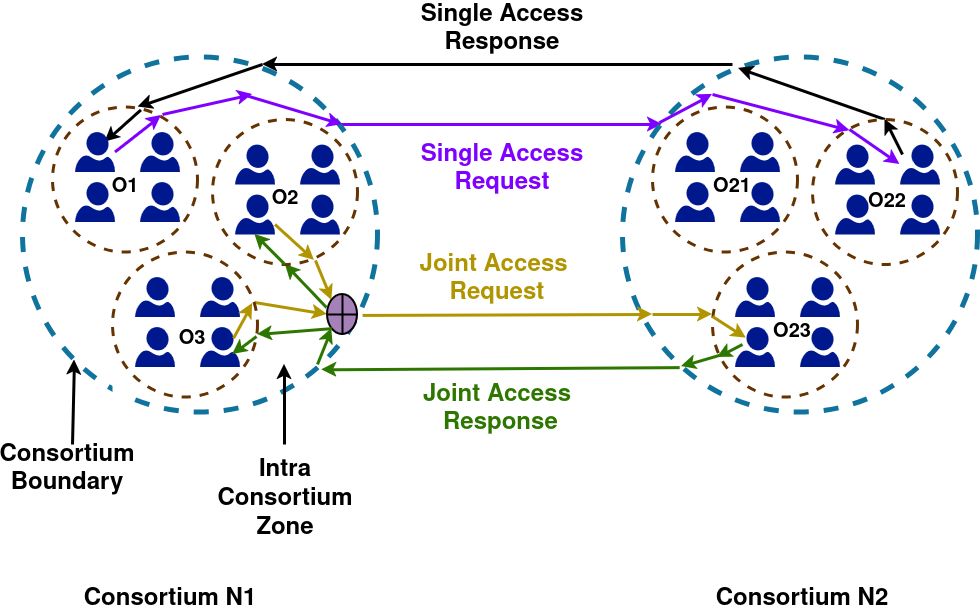}
\caption{Cross-Consortium  Access}
\label{fig:interconsortiumaccess}
\vspace{-0.3cm}
\end{figure}

In cross-blockchain communications, strong access control is essential to protect sensitive data. If a supply chain ledger is compromised, information such as shipping schedules or client details could be exposed.
In the healthcare domain, weak access control could lead to the disclosure of patient records, increasing the risk of identity theft and data misuse. 
Exchanging data and assets between consortiums is complicated, involving distinct networks, each consisting of multiple organizations, as illustrated in Fig. \ref{fig:interconsortiumaccess}. 
Each phase of the interoperation process must be governed by appropriate access control mechanisms and policies. In centralized enterprise systems like ERP and CRM, access is typically managed using \textit{Role-Based Access Control} (RBAC)\cite{akbarfam2023forensiblock}, which relies on user roles and adheres to principles such as least privilege and separation of duties~\cite{samarati2000access,sandhu1996role,zhang2010extended}. For more granular control, \textit{Attribute-Based Access Control} (ABAC)~\cite{wu2023blockchain} is also employed. ABAC uses additional contextual attributes such as time of access, department affiliation, or data sensitivity level to enforce more dynamic and flexible access policies.
Most of these approaches focus on a single blockchain rather than cross-chain communication, and they fail to adequately address access control across enterprise-level consortiums.
To address the limitations, we propose \ours, an access control framework for permissioned blockchain interoperation. 
To address the limitations above, we propose \ours, an access control model for permissioned blockchain interoperation. \ours implements 
endorsement policies based on access request source and access types.
The key contributions of this work, 
are the following. 
\begin{itemize}
\item
\textbf{Access control for the originating network:}
\ours introduces a flexible access control policy that effectively handles varying types of access requirements for the source consortium.
such as individual employees, entire companies, or even joint requests from multiple organizations. 

\item \textbf{Defending Requester Bias:}
This is achieved by
omitting internal details, such as the requesting organization's or employee's identity, thus maintaining privacy. 
\item \textbf{Request unification and privacy:}
It is important to present each inter-consortium message as originating from the consortium as a whole. This is achieved by consolidating outgoing requests while omitting internal details, such as the requesting organization's or actor's identity, thus maintaining privacy. 
\end{itemize}

The proposed framework builds on a preliminary version of a cross-chain access control method that we have introduced in our earlier work \cite{sengupta2025interacct}. 
We test the prototype implementation based on Hyperledger Fabric with Hyperledger Cacti through extensive experiments. 

%% file: Main_Matter/relatedwork.tex
\label{sec:relatedwork}
Recent research has explored various blockchain interoperability approaches. 
Atomic swaps \cite{herlihy2018atomic, reninteroperability} facilitate cross-chain asset transfers
but they face challenges like locking issues. Han et al. \cite{han2023vm} proposed VMs for unified smart contract verifications, which improve functionality but suffer from scalability limitations. Side-chain solutions \cite{back2014enabling, herlihy2018atomic, jin2018towards, kwon2019cosmos, wood2016polkadot} rely on auxiliary chains to handle tasks independently.
Notary schemes, relying on trusted nodes \cite{scheid2021notary}, introduce risks of third-party bias. 
In contrast, blockchain agonistic relay-based inter-blockchain-connector framework is used in Hyperledger Cacti \cite{cactis}. On the other hand, provenance architectures such as global views \cite{belchior2024bungee}, healthcare EHR systems \cite{margheri2020decentralised}, and collaborative system solutions \cite{han2021vassago, pan2023data, augusto2024multi, ctsa2023} are used to ensure exchange of blockchain states with endorsement proofs. 

On the other side, various access control models have evolved over the years, addressing security challenges. Role-Based Access Control (RBAC) \cite{akbarfam2023forensiblock, samarati2000access, sandhu1996role}, a widely used model, adheres to principles like least privilege and duty separation. However, RBAC struggles with granular and dynamic access requirements. Attribute-Based Access Control (ABAC) \cite{yuanabac, hu2013guide} overcomes these limitations by associating policies with user and resource attributes, enabling dynamic and context-aware decisions. In blockchain contexts, decentralized access control is vital. Rouhani et al. \cite{rouhani2021distributed} proposed a distributed ABAC system using permissioned blockchains and chaincode. 
Several decentralized access control approaches \cite{saidi2022dsmac, belchior2020ssibac, braun2023sissi} advocated SSI-based access control frameworks for authentication and authorization.
Si et al. \cite{si2024cross} proposed a cross-chain access scheme for access policies and attributes using relay nodes and homomorphic threshold encryption. Gai et al. \cite{gai2023blockchain} proposed RBAC protocol with multi-signature for cross-organizational data exchange.
These existing systems are often siloed and do not address the inter-consortium access control mechanism for blockchain interoperation.

%% file: Main_Matter/methodology.tex
\begin{figure}[!ht]
\centering 
\includegraphics[width=\linewidth]{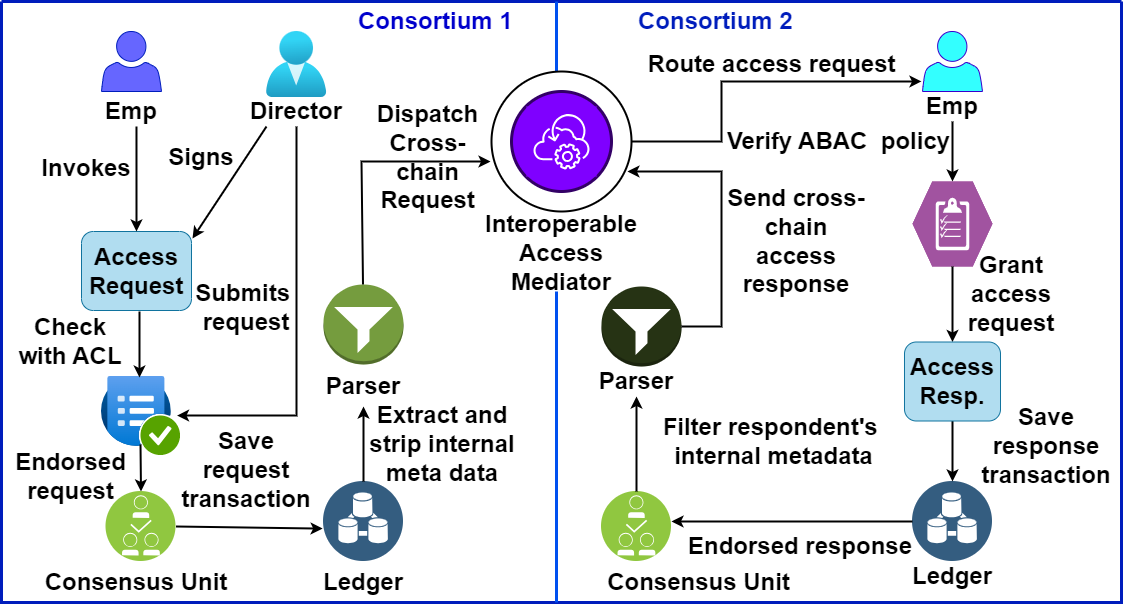}
\caption{\ours Process Flow Diagram}
\label{fig:arch}
\end{figure}

\label{sec:methodology}
The \ours framework is comprised of the following components: (A) \textit{Forming Access Request}, (B) \textit{Extracting and Saving Request Details}, (C) \textit{Interoperable Access Request-Response Data Transfer}, (D)
 \textit{Access Control in Responding Consortium}, and (E) \textit{Encrypting and Routing Access Response}. 
 The overall workflow is depicted in Fig. \ref{fig:arch}. 
Typically, an access request originates from an individual actor within an organization. Using the \textit{intra-organization ACL} policy, \ours verifies whether the requester has the necessary permissions to initiate such a request, after which the signature of the organization’s director is affixed. However, if the director (i.e., the head of the organization) directly initiates the request, no additional organizational-level endorsement is required.
In case of \textit{multilateral} requests, endorsements from the directors of the required organizations are verified.
 In the next step, the request is evaluated against the \textit{consortium ACL} policy, which determines whether the required endorsements from a sufficient number of organizations are present to initiate the cross-blockchain request.
Then the access request ($\mathcal{R}$) is processed to strip internal metadata 
and an access request transaction $T_{\text{acc\_req}}$ is invoked by the consortium ($\mathcal{N}_{1}$). It is
transmitted to foreign blockchain ($\mathcal{N}_{2}$) via an interoperability framework, where it undergoes \textit{cross-consortium ACL} verification  and processing to generate a response. Internal metadata of the responder is filtered out 
and the access response is encrypted. Next a response transaction $T_{\text{acc\_resp}}$ is generated 
forwarded to $\mathcal{N}_{1}$ via the interoperability framework. Finally, the response is routed back to the original requester $\mathcal{E}_{jk}$ in $\mathcal{N}_{1}$.

We detail the components of our proposed framework in the below sub-sections.

\subsection{Forming and Validating Access Request}
\label{subsec:accessrequestmodel} 
When a cross-chain access request transaction $T_{\text{acc\_req}}$ is invoked, it is first signed by the actor $\mathcal{E}_{jk}$. Next, the director of the organization $\mathcal{\psi}_{\text{director}_{\mathcal{O}_{ij}}}$, verifies it as per the $\mathsf{Intra\text{-}Org\ ACL}(\mathcal{R})$ policy and finally endorses it. The Algorithm \ref{algo:1} describes how the access requests are formed, endorsed and validated against the access control polices.
For \textit{unilateral} requests, once an actor $\mathcal{E}_{jk}$ forms and signs an access request as $\mathcal{\psi}_{\mathcal{E}_{jk}}(\mathcal{R})$, the organization director’s signature $\mathcal{\psi}_{\text{director }_{\mathcal{O}_{ij}}}(\mathcal{R})$ is sought to represent the organization-level endorsement $\mathcal{\psi}_{\mathcal{O}_{ij}}(\mathcal{R})$. 
The director's endorsement is considered as an organization's endorsement, $\psi_{\text{director}_{\mathcal{O}_{ij}}}(\mathcal{R}) = \psi_{\mathcal{O}_{ij}}(\mathcal{R})$ (refer steps 1-6 in Algorithm \ref{algo:1}).
In the case of \textit{multilateral} requests, where $S(\mathcal{R})=[\mathcal{E}_{kp}, \mathcal{E}_{kp+1}, \dots, \mathcal{E}_{kq}]$, the request $\mathcal{R}$ must be endorsed collectively by multiple organizations ($\mathcal{E}_{kp}, \mathcal{E}_{kp+1}, \dots, \mathcal{E}_{kq}$) within $\mathcal{N}_1$ (refer steps 7-11 in Algorithm \ref{algo:1}), ensuring that no party can single-handedly approve $\mathcal{R}$.
The algorithm verifies that all mandated organizations have provided valid endorsements before approving the request, verifying $V(\psi_{\mathcal{O}_{ip}} \cap 
\dots \cap \psi_{\mathcal{O}_{iq}}(\mathcal{R}))$ is true or not, as part of the \textit{consortium ACL} policy $\mathsf{Consortium\ ACL}(\mathcal{R})$ (refer steps 12-24 in Algorithm \ref{algo:1}). 
The \textit{consortium ACL} validation is implemented through smart contract logic, and the corresponding ACL policies are pre-configured in the ledger, defining which organizations are permitted to initiate \textit{unilateral} requests and which combinations of organizations can jointly initiate \textit{multilateral} requests.
If all the consortium endorsement rules are met and $V_{\mathcal{O}_{ij}}(\psi)$ evaluates to true, the access request $\mathcal{R}$, along with its attribute values, is committed to the ledger of $\mathcal{N}_1$ with endorsement details $\mathcal{\psi}_{\mathcal{O}_{ij}}$.
If the \textit{ConsortiumACLValidation} is not satisfied within the stipulated time frame, the request $\mathcal{R}$ is marked as invalid and forfeited. 

\subsection{Extracting and Saving Request Details}
\label{subsec:storingaccessrequest}
Directly transmitting an access request $\mathcal{R}$ as it is to a foreign network $\mathcal{N}_2$ may expose the internal details of the requester and the organizational structure of consortium $\mathcal{N}_1$. A foreign consortium could exploit this metadata to infer sensitive roles, hierarchies, or access patterns, and take malicious advantage of this information. To mitigate this, after endorsement, the signed request $\mathcal{\psi}_{\mathcal{O}_{ij}}||\mathcal{R}$ is parsed for extracting critical details such as the original requester's identity, access request reference, node information, and organization details. These details are then recorded in a smart contract of the source network ledger. This process helps in routing the response back to the original requester later. The resulting request therefore represents consortium $\mathcal{N}_1$ as a collective entity as the source, rather than exposing its internal participants.
\input{Main_Matter/algo}
\subsection{Interoperable Access Request-Response Data Transfer}
\label{subsec:iteroperablereqresponsetransfer}
Once the access request transaction $T_{\text{acc\_req}}$ is transmitted by the source consortium $\mathcal{N}_1$, or the access response $T_{\text{acc\_resp}}$ is dispatched by the destination consortium $\mathcal{N}_2$, an interoperability tool is needed to facilitate the data exchange.
The \ours approach incorporates a permissioned blockchain interoperability framework that employs drivers, relays, and attestation-based proof mechanisms to establish cross-chain trust, as initially proposed by Abebe et al. \cite{abebe2019enabling}. This protocol has been further refined in the Hyperledger Cacti project \cite{cactis}. 
When transferring $T_{\text{acc\_req}}$ and $T_{\text{acc\_resp}}$ between $\mathcal{N}_1$ and $\mathcal{N}_2$, the interoperability smart contract first initiates the cross-chain transaction. It prepares the data and associates the proof of consensus on the data as attestations using digital signatures of the consensus participants. The driver component converts the prepared data into a format compatible with any consortium and transmits it to the corresponding relay. The relay, in turn, forwards the data to the driver module of the destination network, which delivers it to the interoperability smart contract on the destination ledger for further processing. 
If either the access request transaction $T_{\text{acc\_req}}$ or the corresponding access response transaction $T_{\text{acc\_resp}}$ does not reach the destination consortium $\mathcal{N}_2$ or the source consortium $\mathcal{N}_1$, respectively, within a stipulated time limit, both the request and the response are marked as invalid in the ledgers of both consortiums.
\subsection{Access Control in Responding Consortium}
\label{subsec:accesscontrolrespondingconsortium}
The interoperability framework Cacti \cite{cactis} provides the transport and attestation proof for identification for cross-consortium data \cite{cactisdatasharingproof}, but does not include a native authentication mechanism to fully verify the legitimacy of the request.
So, \ours ensures that, upon the arrival of the access request transaction $T_{\text{acc\_req}}$ at the destination consortium $\mathcal{N}_2$, the verification process begins by authenticating its origin through validation of $\mathcal{N}_1$'s public key.
Next, \textit{Cross-consortium ACL} evaluates the validation rules for processing access requests that arrive from another consortium. The policy checks the minimum number of organizations in $\mathcal{N}_1$ that must endorse the access request by determining whether $\mathcal{F}$ evaluates to \textit{true}.
Hence, the request is checked for the necessary endorsements as stipulated by $\mathcal{N}_2$'s access control policy by verifying $\mathsf{Eval}\left( \mathsf{Expr}(\mathcal{P}), \mathcal{\psi}_{\mathcal{N}_{1}}\right))$. Similar to $\mathcal{N}_1$, an ACL is defined in $\mathcal{N}_2$ to manage permissions. After verifying the source and legitimacy, the attributes mentioned in the external request are assessed against $\mathcal{N}_2$'s ACL, which incorporates ABAC \cite{yuanabac}. This entire verification process is performed at the smart-contract level by validating the endorsements and cross-checking the corresponding ACL rules.
It evaluates the identifier attributes, resource category, and type of attached endorsements. 
\subsection{Encrypting and Routing Access Response}
\label{subsec:encryptrouteaccessresponse}
Apart from sending the access response, encrypting the response to maintain its confidentiality is also critical.
When an access request is initiated, a temporary encryption key is generated by the requester. The responding consortium uses this key to encrypt the response payload. Thus, even when the response reaches the requesting network, participants other than the original requester cannot decrypt it.
For \textit{multilateral} access requests involving multiple requesters, we employ Threshold Encryption \cite{si2024cross}. Threshold encryption is a cryptographic technique in which a secret key is divided into multiple shares and distributed among several parties such that only a predefined subset (threshold) of them can collaboratively reconstruct the decryption key. In our framework, this ensures that no single requester can decrypt the access response independently; the data can only be revealed when a sufficient number of authorized requesters jointly participate in the decryption process, thereby preserving confidentiality and enforcing joint custodianship of the shared resource.
An encryption key is generated and attached to the request payload, while the shares of the decryption key are distributed across all the requesters. Depending on the consortium policy, at least $t$ out of $n$ requesters need to cooperate to decrypt the response.
Before consortium $\mathcal{N}_1$ sends the payload along with the encryption keys to $\mathcal{N}_2$, it is encrypted using $\mathcal{N}_2$'s public key. This ensures secure transmission of data across consortium boundaries, as we formally prove next.
 

%% file: Main_Matter/algo.tex
\RestyleAlgo{ruled,linesnumbered}
\SetKwComment{Comment}{/* }{ */}
\SetKwInput{KwInput}{Access Request Formation}
\SetKwFunction{AccessValidation}{\textbf{ConsortiumACLValidation}}

\begin{algorithm}[hbt!]
\caption{\textit{\textbf{Consortium ACL Validation}}}
\label{algo:1}
\footnotesize

\KwInput{}  

\If{Request Type == Individual Request}{
    Actor \( \mathcal{E}_{jk} \) generates a \textit{unilateral} request \( \mathcal{R} \)\;
    \( S(\mathcal{R}) \gets \mathcal{E}_{jk} \);  /* Set request source of the request */\\ 
    \( \mathcal{\psi}_{\mathcal{E}_{jk}}(\mathcal{R}) \gets \text{Sign}(\mathcal{E}_{jk}(\mathcal{R})) \); /* Actor signs the request */\\
    Director verifies $\mathcal{R}$ against $\mathsf{Intra\text{-}Org\ ACL}$\;
    \( \mathcal{\psi}_{\mathcal{O}_{ij}}(\mathcal{R}) \gets 
    \mathcal{\psi}_{\mathcal{E}_{jk} = \text{director}}(\mathcal{R} || 
    \mathcal{\psi}_{\mathcal{E}_{jk}}(\mathcal{R})) \);  /* Director provides org-level endorsement*/\;
}
\ElseIf{Request Type == \textit{multilateral} Request}{
    Actor \( \mathcal{E}_{kp} \), \(\mathcal{E}_{kp+1}\), \dots, \( \mathcal{E}_{kq} \) generate an access request \( \mathcal{R} \)\; 
    \( S(\mathcal{R}) \gets [\mathcal{E}_{kp}, \mathcal{E}_{kp+1}, \dots,\mathcal{E}_{kq}] \)\;
    \( \mathcal{\psi}_{\mathcal{E}_{kp}\cup \mathcal{E}_{kp+1}\cup \dots \mathcal{E}_{kq}}(\mathcal{R}) \gets 
    \text{Sign}_{(\mathcal{E}_{kp}\cup \mathcal{E}_{kp+1}\cup \dots \mathcal{E}_{kq})} (\mathcal{R}) \)\;
    \( \mathcal{\psi}_{\mathcal{O}_{ip}\cup \dots \cup \mathcal{O}_{iq}}(\mathcal{R}) \gets 
    \mathcal{\psi}_{(\mathcal{E}_{kp} = \text{director}\cup \mathcal{E}_{kp+1} \dots \mathcal{E}_{kq} = \text{director}) }
    (\mathcal{R} || \mathcal{\psi}_{\mathcal{E}_{kp}\cup \mathcal{E}_{kp+1}\cup \dots \mathcal{E}_{kq}} (\mathcal{R})) \)\;
}
\SetKwProg{Fn}{Function}{:}{}
\Fn{\AccessValidation{\( \mathcal{R}, S(\mathcal{R})\)}}{
/* For \textit{unilateral} requester's request */ \\
    \uIf{\( S(\mathcal{R}) == 
         \mathcal{E}_{jk} \)}{
       /* Return True if organization level endorsement is valid */\\
        \If{\( \text{Validate}(\mathcal{\psi}_{\mathcal{O}_{ij}}(\mathcal{R})) == \mathbb{T}\text{rue} \)}{
            \Return \( \mathbb{T}\text{rue} \)\; 
        }
    }
     /* For \textit{multilateral} requests */\\
    \uElseIf{\( S(\mathcal{R}) == [\mathcal{E}_{kp}, \mathcal{E}_{kp+1}, \dots, \mathcal{E}_{kq}] \)}{
    /* Verify endorsement of each requester's organizations */\\
   \ForEach{\( \mathcal{O}_{l} \in [\mathcal{O}_{ip}, \mathcal{O}_{ip+1}, \dots, \mathcal{O}_{iq}] \)}{
    \If{\( \text{Validate}(\mathcal{\psi}_{\mathcal{O}_{l}}(\mathcal{R})) \neq \mathbb{T}\text{rue} \)}{
        \Return \( \mathbb{F}\text{alse} \)\; 
    }
}
\Return \( \mathbb{T}\text{rue} \)\;
}
}
\end{algorithm}

%% file: Main_Matter/analysis.tex
\label{sec:analysis}
In addition, \ours prevents three major categories of attacks: defending insider threats, preventing carry-forward of internal information, and guarding against sniffing attacks. We further analyze how \ours mitigates these threats below. 

\subsection{Defending Insider Threat}  
\label{subsec:attacksdefended}
In \ours every access request $\mathcal{R}$ is subject to a multi-stage access control process. The \textit{intra-organization ACL} blocks unauthorized requests that have not been validated and endorsed by the ${\text{director}_{{\mathcal O}_{ij}}}$ of the organization $\mathcal O_{ij}$. \textit{Consortium ACL} ensures that no access request from consortium $\mathcal{N}_{1}$ can be transmitted to a foreign consortium $\mathcal{N}_{2}$ unless it carries the requisite consortium-level endorsements, thereby preventing bypass of consortium approval. Finally, the \textit{cross-consortium ACL} ensures that access requests failing to comply with agreed cross-chain access norms are also rejected. In the following we prove \ours successfully deters insiders threats.

\begin{IEEEproof}
Assume that an insider participant in consortium $\mathcal{N}_{1}$ attempts to gain access to data/resources of $\mathcal{N}_{2}$ without valid authorization.
\par\textit{Case 1:}  
An actor $\mathcal{E}_{jk} \in \mathcal{O}_{1j}$ of $\mathcal{N}_{1}$ tries to send a direct request $\mathcal{R}$ to $\mathcal{N}_{2}$ without the director’s endorsement. The $\mathsf{Intra\text{-}Org\ ACL}(\mathcal{R})$ requires that $\mathcal{\psi}_{\text{director}_{\mathcal{O}_{ij}}}(\mathcal{R})$ be attached. Since this endorsement is absent, the request is rejected locally in $\mathcal{N}_{1}$. This contradicts the assumption that the actor can bypass \textit{intra-organization ACL} and succeed to gain external access.  
\par\textit{Case 2:}  
Suppose the director of an organization within $\mathcal{N}_{1}$ does not obey \textit{consortium} level endorsement norms and maliciously attempts to invoke $\mathcal{R}$ directly towards consortium $\mathcal{N}_{2}$. The $\mathsf{ConsortiumACLValidation}(\mathcal{R})$ (refer Algorithm~\ref{algo:1}) requires that signatures from the required organizations in $\mathcal{N}_{1}$ to be present. Since the necessary collective endorsements are missing, the request is rejected at the consortium boundary. This contradicts the assumption.  

Hence, \ours prevents unauthorized access attempts of malicious insiders.
\end{IEEEproof}

\subsection{Preventing Carry-forward of Internal Information}  
If confidential internal metadata of the source consortium is exposed in inter-consortium data exchange, malicious participants could reverse-engineer these.
\ours defends against this by stripping internal metadata $\mathcal{R}_\epsilon$ from each access request, transforming it into $\mathcal{R} = \mathcal{R} - \mathcal{R}_\epsilon$ before transmission(refer to Sec.\ref{subsec:storingaccessrequest}).  

\subsection{Guarding against Sniffing Attack}  
In \ours, all cross-consortium responses are encrypted using the temporary key or key shares generated at request time. 
Thus, even if data is tried to be intercepted, it cannot be accessed by unauthorized organizations (refer to Sec.\ref{subsec:encryptrouteaccessresponse}).

\begin{IEEEproof}
Consider a malicious outsider tries to sniff the network for retrieving access-request and/or access-response details between two interoperating permissioned networks $\mathcal{N}_{1}$ and $\mathcal{N}_{2}$.

\textit{Case 1:}
Assume that the adversary intercepts an encrypted access response and retrieves the response payload for a \textit{unilateral} access request transmitted from $\mathcal{N}_{2}$ to $\mathcal{N}_{1}$. In \ours, each \textit{unilateral} access response is encrypted using the requester’s temporary private key, which is not transmitted or exposed over the network and remains solely under the custody of the requester. Since this private key is not shared or included in the ciphertext, successful retrieval by the adversary would contradict the assumed security of the public-key cryptographic principles. Therefore, the adversary cannot retrieve or reconstruct the actual response content from the intercepted ciphertext.

\textit{Case 2:}
Assume that the adversary successfully extracts plain-text from an encrypted access response  for a \textit{multilateral} access request transmitted from $\mathcal{N}_{2}$ to $\mathcal{N}_{1}$. In \ours, each \textit{multilateral} request employs a fresh $t$-of-$n$ threshold encryption scheme, where fewer than $t$ key shares yield no useful information about the decryption key. Since the shares are distributed only among the legitimate requesters and never reused across transactions, the adversary, lacks the threshold number of key shares. Hence they cannot decrypt or fabricate a valid response transaction.

Therefore, \ours prevents sniffing attacks by design.
\end{IEEEproof}

\subsection{Safety and Liveness Guarantees}
\label{subsec:safety_liveness}
The \ours architecture ensures several critical security guarantees by combining multi-faceted ACL validation, threshold encryption, secure response routing, and immutable recording of access request--response transactions on consortium blockchains. In the following, we discuss the safety and liveness guarantees that \ours provides for inter-consortium access control.

\subsubsection{Safety}
In \ours an access request is only allowed if it passes (a) signature verification and (b) ACL validation. This validation is enforced in three phases of the interoperability flow:  
(i) within an organization, where the director validates the \textit{intra-organization ACL} before signing a request;  
(ii) within a consortium, where the \textit{ConsortiumACLValidation} validates endorsements $\mathcal{\psi}_{\mathcal{O}_{ij}}(\mathcal{R})$ from the originating organization $\mathcal{O}_{ij}$ for unilateral requests, and for \textit{multilateral} requests (refer to Algorithm~\ref{algo:1}). In  \textit{multilateral} requests, endorsements $\bigcap_{l=p}^{q}\mathcal{\psi}_{\mathcal{O}_{il}}(\mathcal{R})$ from multiple organizations $(\mathcal{O}_{ip}, \dots, \mathcal{O}_{iq})$ are scrutinized;
(iii) across consortium boundaries, where $\mathsf{Cross\text{-}consortium\ ACL}(\mathcal{R})$ ensures that the destination consortium verifies the attributes of $\mathcal{R}$. Apart from these, the encryption technique is used to secure the custody of the access responses. These rigorous validations collectively preserve safety against unauthorized access and malicious faults. 
We further elaborate on the individual safety properties:
\begin{itemize}
\item \textbf{Confidentiality:} Both the request payload, as well as the response data/asset are encrypted to ensure confidentiality. For \textit{multilateral} requests from more than one stakeholder, the decryption key is split into shares using threshold cryptography (refer to Section~\ref{subsec:encryptrouteaccessresponse}). The access response is encrypted so that only the originator can possess the response. Routing of encrypted access response and internal metadata filtering (refer to Section~\ref{subsec:storingaccessrequest}) collectively ensure that any sensitive information remains confidential.
\item \textbf{Integrity:} Integrity is maintained by digital signatures at every stage of the workflow, from actors within an organization, to directors, to the responding consortium. Access requests and responses, along with endorsements, are committed as $T_{\text{acc\_req}}$ and $T_{\text{acc\_resp}}$ in the ledgers $\mathcal{L}_{\mathcal{N}_1}$ and $\mathcal{L}_{\mathcal{N}_2}$ (refer to Section~\ref{subsec:storingaccessrequest}). These records are stored immutably, making them tamper-proof.

\item \textbf{Authenticity:} Each request $\mathcal{R}$ is signed by the requester and the designated endorsers. The \texttt{ConsortiumACLValidation} sub-routine (refer to Algorithm~\ref{algo:1} in Section~\ref{subsec:accessrequestmodel}) ensures that only legitimate participants can originate access requests.

\item \textbf{Non-repudiation:} Once transactions $T_{\text{acc\_req}}$ and $T_{\text{acc\_resp}}$ are recorded on the ledgers with attached endorsements (refer to Section~\ref{subsec:storingaccessrequest}), neither consortium can deny their involvement. Immutable storage of endorsements and responses ensures that all participants remain accountable.
\end{itemize}


\subsubsection{Liveness}
Liveness denotes the system’s ability to make continual progress despite faults or delays. In \ours, this property is preserved through timeout-based validation and the underlying consensus of permissioned blockchains.

\begin{itemize}
\item \textbf{Transaction Handling:} Failures or unreasonable delays in completing an access request-response are handled by stipulated time limits, ensuring that the system continues without indefinite blocking. If they do not reach beyond the limit, they are marked as invalid.

\item \textbf{Consensus Progress:} The liveness of request and response execution depends on the consensus underlying blockchain platform (e.g., Hyperledger Fabric). As long as a majority of peers remain honest and active, consensus guarantees that valid transactions $T_{\text{acc\_req}}$ and $T_{\text{acc\_resp}}$ are eventually committed.

\end{itemize}

Collectively, these validation mechanisms ensure that \ours maintains safety while guaranteeing liveness so that legitimate access requests are eventually processed and responded to.

%% file: Main_Matter/implementation.tex
\label{sec:implemenation}
We have implemented \ours using Hyperledger Fabric as the permissioned blockchain network and used Hyperledger Cacti \cite{cactis} for data interoperation.
The Hyperledger Cacti \cite{cactis} based interoperability framework enables cross-chain communication of access request-response via relays, drivers, and interoperability chaincode in the data interoperability layer. 
We have registered and enrolled the roles for different employees and directors of the organizations, and the associated attributes at the CA certificate level using native Hyperledger Fabric ABAC offering \cite{hlfabac}.
The core  components of \ours are detailed below.
\par \textbf{Access Request Formation:}
The process begins with a participant from a consortium generating a cross-chain access request.
The private and public keys of the request generator are stored in .pem files, and the signing is performed using the Elliptic Curve Digital Signature Algorithm (ECDSA) \cite{ellipcticcurve_johnson1999elliptic} with SHA-256 \cite{sha_nist} hashing.
To enforce the rules governing request submission, an Access Control List (ACL) is created and maintained by \textit{ManageAccessContol} chaincode. 
The participant's request generation is handled through our \textit{SubmitAccessRequest} chaincode. The intra-organization endorsements of the request is verified by the \textit{verifySignature} function within the chaincode. ACL verification by \textit{consortiumACLValidation} subroutine
verifies whether the request has been authenticated by the participant and includes the necessary co-signature reference details. 
\par \textbf{Saving and Filtering Access Request:}
A unique request ID is generated for each access request by a participant of $\mathcal{N}_1$, and the request details are recorded as a transaction in the ledger of $\mathcal{N}_1$, after satisfying the endorsement policy of $\mathcal{N}_1$. 
The majority endorsement rule is applied at the chaincode level, requiring approval from more than two-thirds of the participants of $\mathcal{N}_1$ before the access request transaction 
can be committed to the ledger. 
Once committed, the request is stored in a private data collection (PDC) \cite{hlfpdc}
repository. PDC stores sensitive data privately among the authorized parties while maintaining a hash of the data on the ledger for integrity verification. We deploy \textit{saveRequestPDC} subroutine to privately save the request based on our permission configuration setup defined in \textit{MaintainCollectionConfig} config file.
The repository stores the following information: requester name, request ID, peer/node ID, organization ID, object requested, etc.
Next, personal information, such as the requester's organization ID, internal role, peer ID, etc. is filtered out from the original access request to prevent potential bias from the responder consortium, $\mathcal{N}_2$. 
Consortium-level tagging is appended to the request.
The interoperability framework \cite{cactis} is then invoked to transfer the request to $\mathcal{N}_2$.  
\par \textbf{Receiving Incoming Access Request:}
We utilize Cacti \cite{cactis} based interoperability framework on both networks $\mathcal{N}_1$ and $\mathcal{N}_2$. The interoperability service routine is enabled for both outward and inward data/asset transfer through the relay mechanism.
The interoperable relay, serving as an intermediary service, invokes the gRPC\footnote{\url{https://grpc.io/} (Accessed: \today)} based lightweight message exchange protocol to transmit the access response payload from $\mathcal{N}_2$ to $\mathcal{N}_1$.
Upon receipt of the incoming access request,
$\mathcal{N}_2$ extracts the relevant attributes including source network ID, destination network ID, request ID, endorsement details, the object requested, etc.,  using \textit{extractAccessReuest} service of our \textit{ManageInteropRequest} chaincode. These attributes are subjected to verification against the cross-consortium access control policies of $\mathcal{N}_2$. We ensure these validations using \textit{checkAccessResponder} subroutine.
If the request meets the necessary conditions and is approved, the requested resource is retrieved and the access response transaction
is marked as complete.
\par \textbf{Transfer of Inter-Chain Access Response:}
After the response is created by the participant in the responding consortium, 
we create response transaction 
and record in $\mathcal{N}_2$'s ledger.
Next, we invoke interoperability chaincode to send the 
response from  $\mathcal{N}_2$ to $\mathcal{N}_1$.
Upon receipt of the access response, $\mathcal{N}_1$ processes and verifies the authenticity of received data. Based on the originating request ID, the original request details are fetched from the request store, and a chaincode-based event is invoked to notify the originating party. The originating requester obtains the access response, and the response transaction is marked as complete.

%% file: Main_Matter/results.tex
\label{sec:results}
To evaluate \ours, we have experimented extensively with two Hyperledger Fabric test networks. The testbed is set up using four 
physical machines with Intel i5-4570 CPU, 8 GB RAM, and 64-bit Ubuntu 20.04 LTS environment, connected through Docker Swarm \cite{dockerswarm} based overlay networks.  
\par
\begin{figure}[t]
  \centering
  \begin{minipage}{0.96\linewidth}
    \centering
    \includegraphics[width=\linewidth,, keepaspectratio]{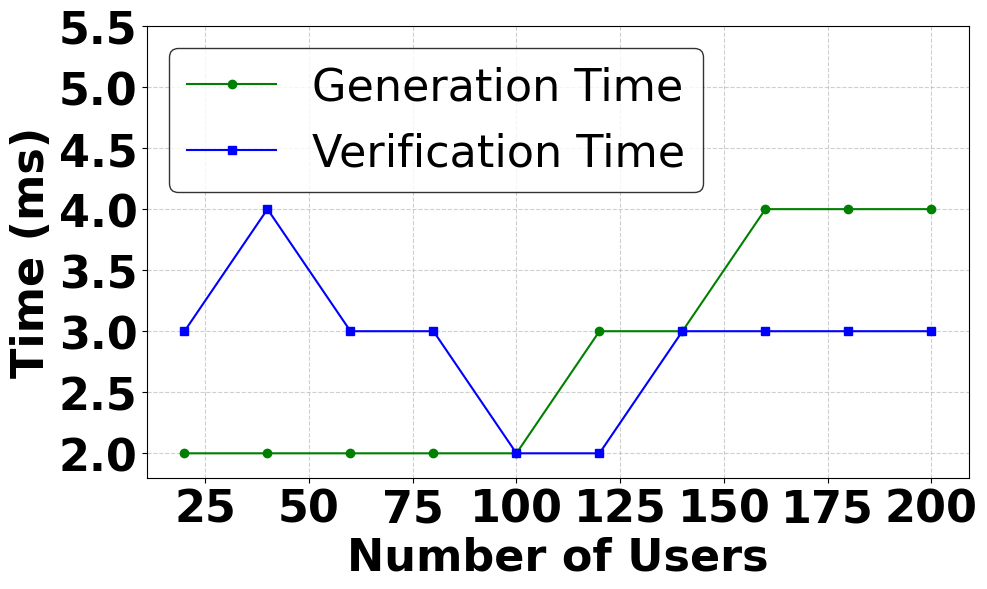}
    \caption{Signature Generation and Verification}
    \label{fig:sign_gen_verify}
  \end{minipage}
  \begin{minipage}{0.96\linewidth}
    \centering
    \includegraphics[width=\linewidth, , keepaspectratio]{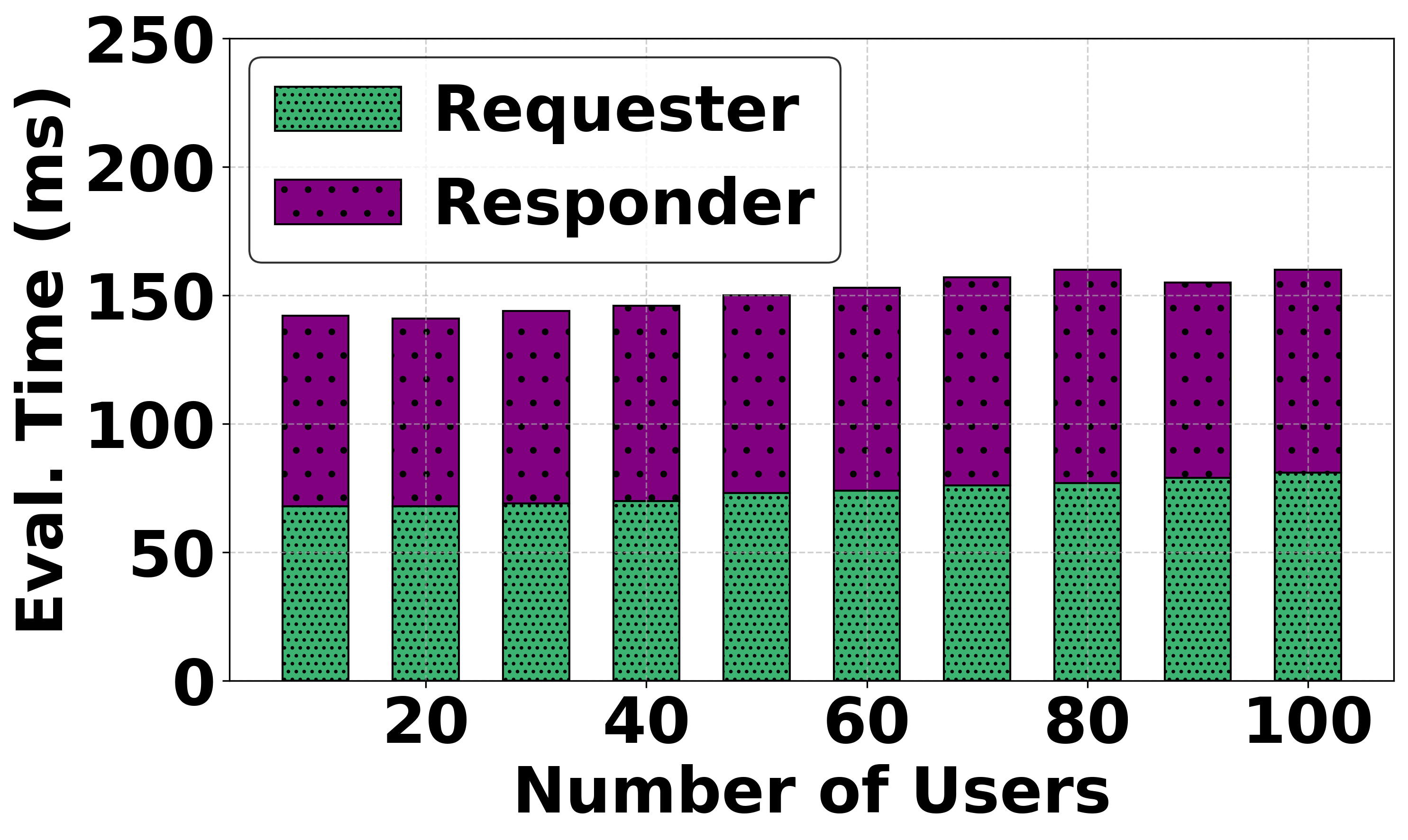}
    \caption{Access Policy Evaluation}
    \label{fig:access_policy_eval}
  \end{minipage}
\end{figure}

\noindent \textbf{Signature Generation and Verification:}
As depicted. Fig. \ref{fig:sign_gen_verify} represents the average time taken to generate the attestation signature accompanying the access request. We test this with a varying set of signature generation requests in terms of director signature, joint-organization-level signatures, etc. The signature generation time remains consistently low at approximately $2$ milliseconds for up to $100$ users. Beyond this point, an increase is observed, with the time reaching up to $4$ milliseconds by around $160$ users, after which it again stabilizes. 
Fig. \ref{fig:sign_gen_verify} also presents the signature verification latency trend with respect to increasing user load. It is observed that the graph stabilizes at $3$ milliseconds from $140$ users onward and remains steady. Both signature generation and verification are efficient. Even under a load of $200$ users, the maximum observed overhead is only 4 milliseconds, and does not contain any significant fluctuations. 

\par\noindent \textbf{Access Policy Evaluation:}  
\ours employs access policy validation at three levels: intra-organization, consortium, and cross-consortium. Fetching the ACL and validating it against each access request is a critical step for enforcing interoperation governance. 
We evaluated the average policy evaluation time at both the source (requester) consortium and the destination (responder) consortium, as shown in Fig. \ref{fig:access_policy_eval}. The results show that requester-side evaluation grows only moderately, from about $67$ milliseconds at $10$ users to $80$ milliseconds at $100$ users. On the responder side, validation remains nearly constant between $74$ milliseconds and $80$  milliseconds across the same load range. Consequently, the combined evaluation time increases slightly from around $141$ milliseconds to $156$ milliseconds as user concurrency increases.  
These results demonstrate that the validation function $\mathcal{P}$ performs steadily, adding only marginal overhead even under higher loads.
\begin{figure}[t]
\centering 
  \begin{minipage}{0.96\linewidth}
    \centering
\includegraphics[width=\linewidth, keepaspectratio]{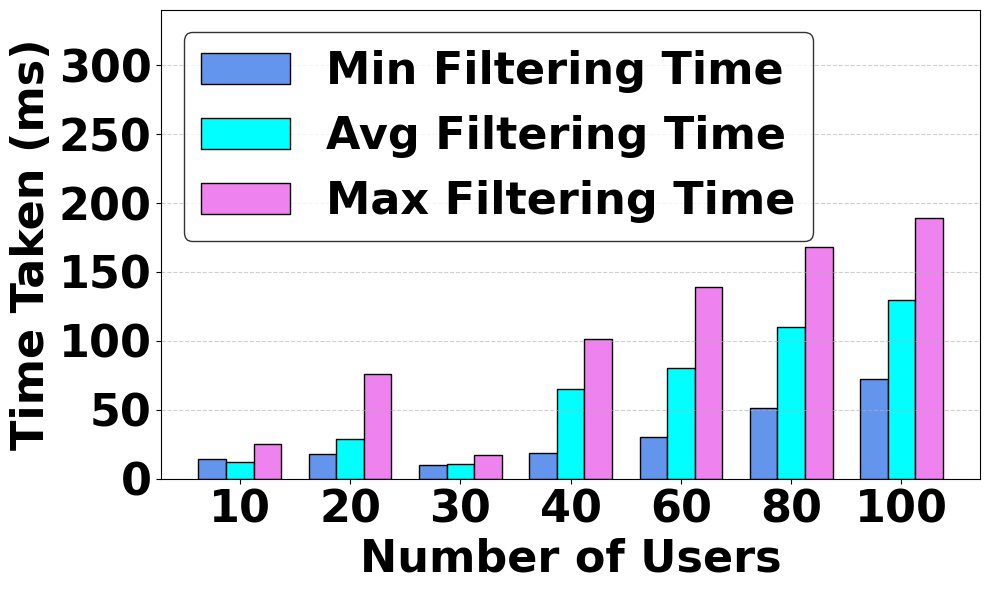}
\caption{Request Filtering}
\label{fig:filter_access_request}
 \end{minipage}
 \begin{minipage}{0.96\linewidth}
  \centering
  \includegraphics[width=\linewidth, keepaspectratio]{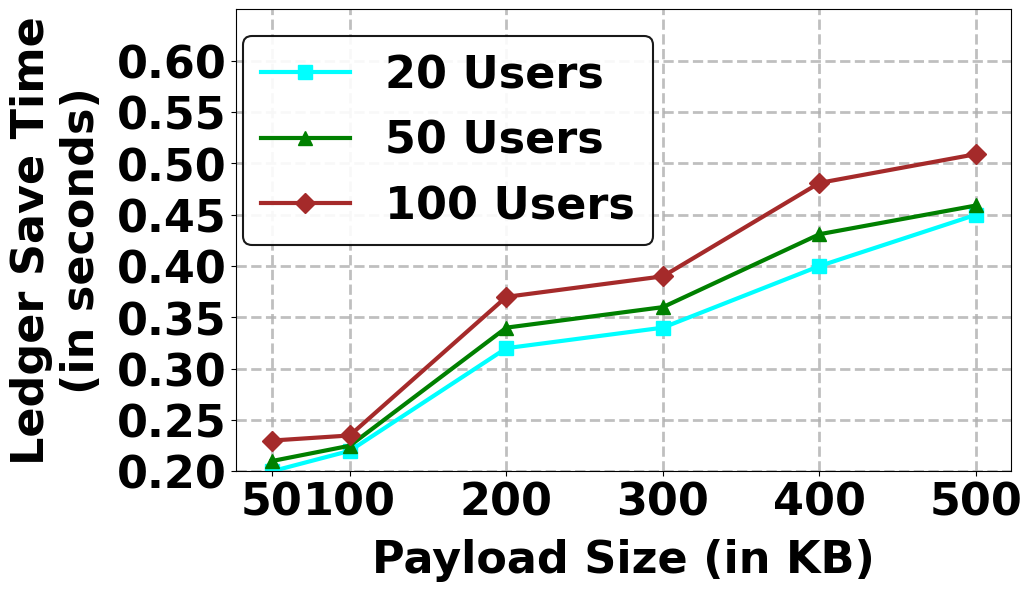}
  \caption{Ledger Save Time}
  \label{fig:saveaccessrequest_requester}
\end{minipage}
\end{figure}
\par\noindent \textbf{Request Filtering:}
Fig. \ref{fig:filter_access_request} illustrates the time required for filtering access requests to remove internal role details. We have tested with a growing number of users (from $10$ to $100$) and noticed that the filtering time completes under $0.2$ seconds. This demonstrates that filtering introduces no performance bottleneck and completes with minimal latency.

To evaluate the ledger commit performance for access requests in Hyperledger Fabric, we analyze how varying payload sizes impact the save time across different user counts, as illustrated in Fig. \ref{fig:saveaccessrequest_requester}. The results demonstrate that as the payload size increases from $50$ KB to $500$ KB, the ledger save time follows linear trend. 

\par\noindent \textbf{Unauthorized Access Validation:}
\begin{figure}[!t]
\centering 
\begin{minipage}{0.96\linewidth}
\centering 
\includegraphics[width=\linewidth, keepaspectratio]{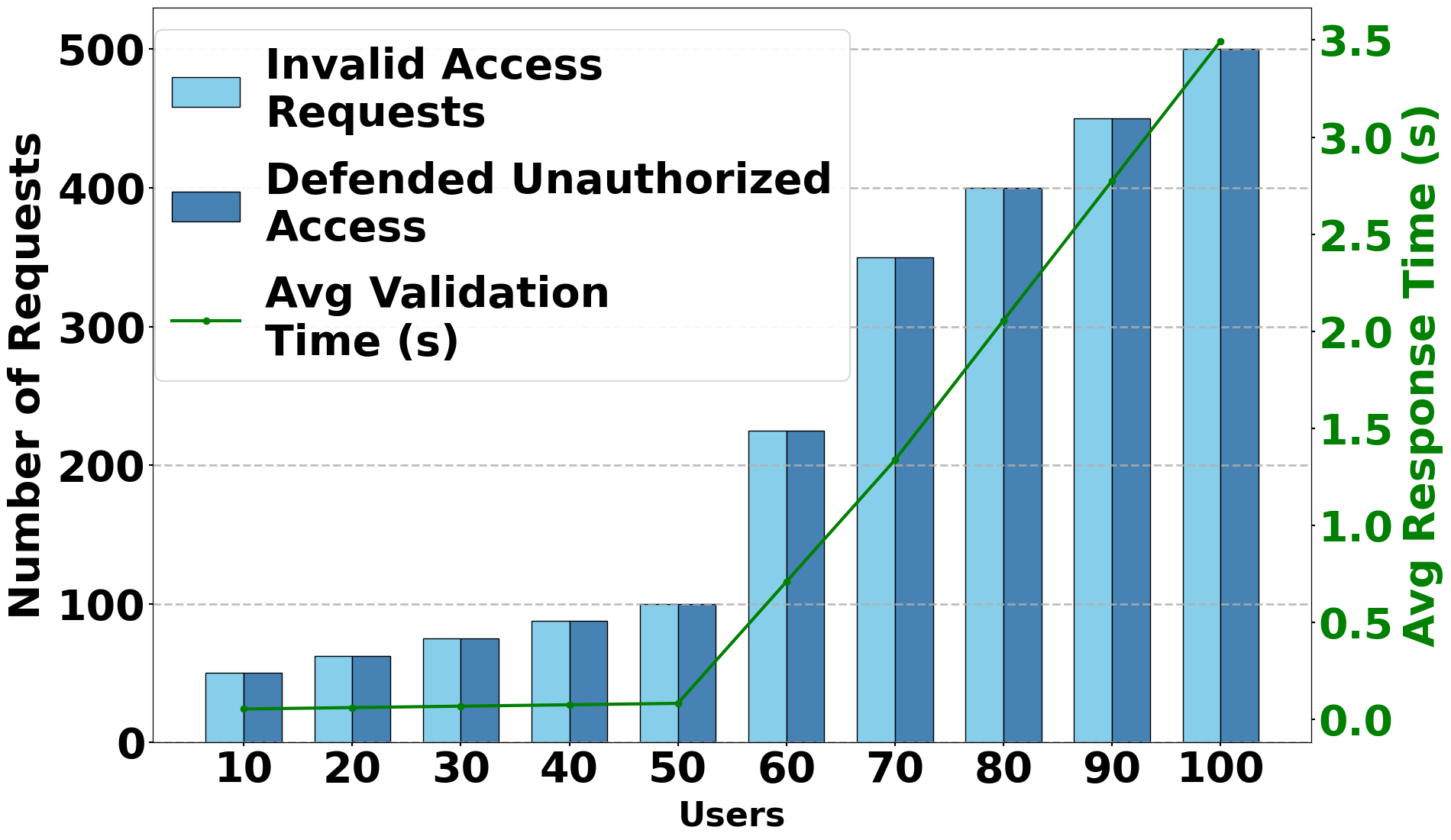}
\caption{Unauthorized Access Validation}
\label{fig:unauthroized_access_validation}
\end{minipage}
\end{figure}
We also assess the impact of unauthorized access based on \ours under high load to verify the effectiveness of access policy validation against invalid or malicious requests, as depicted in Fig. \ref{fig:unauthroized_access_validation}. It is observed that, with the increase in user load from $10$ to $100$ users, all the invalid access requests are handled by the access policy validation without any erroneous access granting. We noticed that the average ACL policy validation time increases under load, following a gradual linear increase in policy evaluation response time due to concurrent validation requests.

%% file: Main_Matter/conclusion_futurework.tex
\label{sec:conclusion_futurework}
We have presented \textit{InterAcct}, a decentralized and secure cross-network access control framework for permissioned blockchain interoperation. The proposed solution introduces a novel attribute-centric access control mechanism that enforces fine-grained authorization across consortium boundaries, while enabling encrypted and verifiable transfer of access responses between networks. \textit{InterAcct} ensures end-to-end data integrity and privacy by combining data filtering, encryption, and multi-level endorsements with cryptographic attestations. In future work, we aim to extend the framework with real-time monitoring and auditing capabilities to facilitate continuous oversight of access requests, approvals, and resource sharing across interconnected consortium networks.
